\documentclass[letter]{aa} 

\usepackage{xspace}
\usepackage{graphicx}
\usepackage{txfonts}
\usepackage{booktabs}
\usepackage{url}
\usepackage{xcolor}
\usepackage[breaklinks, colorlinks, citecolor=blue, linkcolor=blue]{hyperref}
\usepackage{scalerel}
\usepackage{tikz}

\newcommand{\Angstrom}{\AA{}ngstr\"om\xspace}
\newcommand{\mtwo}{MuSCAT2\xspace}
\newcommand{\mthree}{MuSCAT3\xspace}

\newcommand{\pytransit}{\textsc{PyTransit}\xspace}
\newcommand{\ldtk}{\textsc{LDTk}\xspace}

\newcommand{\ud}{\ensuremath{\mathrm{d}}\xspace}

\begin{document} 

\DeclareRobustCommand{\orcidicon}{%
    \begin{tikzpicture}
    \draw[lime, fill=lime] (0,0) 
    circle [radius=0.16] 
    node[white] {{\fontfamily{qag}\selectfont \tiny ID}};
    \draw[white, fill=white] (-0.0625,0.095) 
    circle [radius=0.007];
    \end{tikzpicture}
    \hspace{-2mm}
}

\newcommand*{\tabhead}[1]{\multicolumn{1}{c}{\bfseries #1}}
\newcommand{\mystar}{{\Large {\fontfamily{lmr}\selectfont$\star$}}}
\newcommand{\pds}{PDS~70}

\newcommand{\jgrp}{J. Geophys. Res.-Planets}
\newcommand{\rsos}{Royal Soc. Open Sci.}
\newcommand{\astrolett}{Astron. Lett.}
\newcommand{\psj}{Planet. Sci. J.}
\newcommand{\jatis}{J. Astron. Telesc. Instrum. Syst.}

\newcommand{\mgtwo}{Mg\,II}
\newcommand{\ctwo}{C\,II}
\newcommand{\molh}{H$_2$}
\newcommand{\lyalph}{Ly-$\alpha$}
\newcommand{\fetwo}{Fe\,II}

\newcommand{\lognh}{$\log(N_{\rm H~I})$}
\newcommand{\oi}{[O\,I]}
\newcommand{\suii}{[S\,II]}
\newcommand{\nii}{[N\,II]}
\newcommand{\nai}{Na\,I}
\newcommand{\caii}{Ca\,II}
\newcommand{\mdotacc}{$\cdot{M}_{\rm acc}$}
\newcommand{\mdotwind}{$\cdot{M}_{\rm wind}$}
\newcommand{\hei}{He\,I}
\newcommand{\lyalpha}{Ly-$\alpha$}
\newcommand{\halpha}{H$\alpha$}
\newcommand{\mgi}{Mg\,I}
\newcommand{\sii}{Si\,I}
\newcommand{\water}{H$_2$O}
\newcommand{\methane}{CH$_4$}
\newcommand{\cotwo}{CO$_2$}
\newcommand{\htwo}{H$_2$}


\newcommand{\rosat}{\emph{Rosat}}
\newcommand{\galex}{\emph{GALEX}}
\newcommand{\tess}{\emph{TESS}}
\newcommand{\plato}{\emph{PLATO}}
\newcommand{\gaia}{\emph{Gaia}}
\newcommand{\ktwo}{\emph{K2}}
\newcommand{\jwst}{\emph{JWST}}
\newcommand{\kepler}{\emph{Kepler}}
\newcommand{\corot}{\emph{CoRoT}}
\newcommand{\hipp}{\emph{Hipparcos}}
\newcommand{\spitzer}{\emph{Spizter}}
\newcommand{\herschel}{\emph{Herschel}}
\newcommand{\hst}{\emph{HST}}
\newcommand{\wise}{\emph{WISE}}
\newcommand{\swift}{\emph{Swift}}
\newcommand{\chandra}{\emph{Chandra}}
\newcommand{\xmm}{\emph{XMM-Newton}}

\newcommand{\twa}{TW Hydra}
\newcommand{\bpic}{$\beta$~Pictoris}
\newcommand{\abdor}{AB~Doradus}
\newcommand{\rup}{Ruprecht~147}
\newcommand{\etacha}{$\eta$\,Chamaeleontis}
\newcommand{\usco}{Upper~Sco}
\newcommand{\rhooph}{$\rho$~Oph}

\newcommand{\doar}{DoAr\,25}
\newcommand{\epcha}{EP\,Cha}
\newcommand{\rylup}{RY\,Lup}
\newcommand{\hdtwofour}{HD\,240779}

\newcommand{\sigrv}{$\sigma_{\rm RV}$}

\newcommand{\msunyr}{\rm{M_{\sun} \, yr^{-1}}}
\newcommand{\etal}{\mbox{\rm et al.~}}
\newcommand{\ms}{\mbox{m\,s$^{-1}~$}}
\newcommand{\kms}{\mbox{km\,s$^{-1}~$}}
\newcommand{\ks}{\mbox{km\,s$^{-1}~$}}
\newcommand{\kse}{\mbox{km\,s$^{-1}$}}
\newcommand{\mse}{\mbox{m\,s$^{-1}$}}
\newcommand{\msy}{\mbox{m\,s$^{-1}$\,yr$^{-1}~$}}
\newcommand{\msye}{\mbox{m\,s$^{-1}$\,yr$^{-1}$}}
\newcommand{\msun}{M$_{\odot}$}
\newcommand{\msune}{M$_{\odot}$}
\newcommand{\rsun}{R$_{\odot}$}
\newcommand{\lsun}{L$_{\odot}~$}
\newcommand{\rsune}{R$_{\odot}$}
\newcommand{\mjup}{M$_{\rm JUP}~$}
\newcommand{\mjupe}{M$_{\rm JUP}$}
\newcommand{\msat}{M$_{\rm SAT}~$}
\newcommand{\msate}{M$_{\rm SAT}$}
\newcommand{\mnep}{M$_{\rm NEP}~$}
\newcommand{\mnepe}{M$_{\rm NEP}$}
\newcommand{\mearth}{$M_{\oplus}$}
\newcommand{\mearthe}{$M_{\oplus}$}
\newcommand{\rearth}{$R_{\oplus}$}
\newcommand{\rearthe}{$R_{\oplus}$}
\newcommand{\rjup}{R$_{\rm JUP}~$}
\newcommand{\msinie}{$M_{\rm p} \sin i$}
\newcommand{\vsinie}{$V \sin i$}
\newcommand{\mbsini}{$M_b \sin i~$}
\newcommand{\mcsini}{$M_c \sin i~$}
\newcommand{\mdsini}{$M_d \sin i~$}
\newcommand{\chisq}{$\chi_{\nu}^2$}
\newcommand{\chinu}{$\chi_{\nu}$}
\newcommand{\chinusq}{$\chi_{\nu}^2$}
\newcommand{\arel}{$a_{\rm rel}$}
\newcommand{\feh}{\ensuremath{[\mbox{Fe}/\mbox{H}]}}
\newcommand{\rphk}{\ensuremath{R'_{\mbox{\scriptsize HK}}}}
\newcommand{\lrphk}{\ensuremath{\log{\rphk}}}
\newcommand{\cs}{$\sqrt{\chi^2_{\nu}}$}
\newcommand{\etaearth}{$\mathbf \eta_{\oplus} ~$}
\newcommand{\etaearthe}{$\mathbf \eta_{\oplus}$}
\newcommand{\searth}{$S_{\bigoplus}$}
\newcommand{\mdotyr}{$M_{\odot}$~yr$^{-1}$}
\newcommand{\micron}{$\mu$m}

\newcommand{\sini}{\ensuremath{\sin i}}
\newcommand{\msini}{\ensuremath{M_{\rm p} \sin i}}
\newcommand{\mplsini}{\ensuremath{\mpl\sin i}}
\newcommand{\teff}{\ensuremath{T_{\rm eff}}}
\newcommand{\teq}{\ensuremath{T_{\rm eq}}}
\newcommand{\logg}{\ensuremath{\log{g}}}
\newcommand{\vsini}{\ensuremath{v \sin{i}}}
\newcommand{\ebv}{E($B$-$V$)}

\newcommand{\Kepler}{\emph{Kepler}~}
\newcommand{\Keplere}{\emph{Kepler}}
\newcommand{\blender}{{\tt BLENDER}~}

\newcommand{\kp}{\ensuremath{\mathrm{Kp}}}
\newcommand{\rstar}{\ensuremath{R_\star}}
\newcommand{\mstar}{\ensuremath{M_\star}}
\newcommand{\loggstar}{\ensuremath{\logg_\star}}
\newcommand{\lstar}{\ensuremath{L_\star}}
\newcommand{\astar}{\ensuremath{a_\star}}
\newcommand{\loglstar}{\ensuremath{\log{L_\star}}}
\newcommand{\rhostar}{\ensuremath{\rho_\star}}

\newcommand{\rp}{\ensuremath{R_{\rm p}}}
\newcommand{\rpl}{\ensuremath{r_{\rm P}}~}
\newcommand{\rple}{\ensuremath{r_{\rm P}}}
\newcommand{\mpl}{\ensuremath{m_{\rm P}}~}
\newcommand{\lpl}{\ensuremath{L_{\rm P}}~}
\newcommand{\rhopl}{\ensuremath{\rho_{\rm P}}~}
\newcommand{\loggpl}{\ensuremath{\logg_{\rm P}}~}
\newcommand{\logrpl}{\ensuremath{\log r_{\rm P}}~}
\newcommand{\logrple}{\ensuremath{\log r_{\rm P}}}
\newcommand{\loga}{\ensuremath{\log a}~}
\newcommand{\logmpl}{\ensuremath{\log m_{\rm P}}~}

\newcommand{\fludensunits}{ergs s$^{-1}$ cm$^{-2}$ \AA$^{-1}$}

   \title{Climate Change in Hell:\\ Long-Term Variation in Transits of the Evaporating Planet K2-22b}

   \author{
   E.~Gaidos\inst{\ref{vienna},\ref{hawaii}\thanks{eric.gaidos@univie.ac.at}} \and
   H.~Parviainen\inst{\ref{iull},\ref{iiac}} \and
   E.~Esparza-Borges\inst{\ref{iiac},\ref{iull}}  \and
   A.~Fukui\inst{\ref{ikis},\ref{iiac}}
   K.~Isogai\inst{\ref{iut}, \ref{ioo}} \and
   K.~Kawauchi\inst{\ref{iur}} \and
   J.~de~Leon\inst{\ref{iut}} \and
   M.~Mori\inst{\ref{ica},\ref{naoj}} \and
   F.~Murgas\inst{\ref{iiac},\ref{iull}}  \and
   N.~Narita\inst{\ref{ikis},\ref{iiac},\ref{ica}} \and
   E.~Palle\inst{\ref{iiac},\ref{iull}} \and
   N.~Watanabe\inst{\ref{iut}} 
   }

    \institute{Institute for Astrophysics, University of Vienna, 1180 Vienna, Austria\label{vienna} \and
        Department of Earth Sciences, University of Hawai'i at M\"{a}noa, Honolulu, Hawai'i 96822 USA\label{hawaii} \and
        Departamento Astrof\'isica, Universidad de La Laguna (ULL), E-38206 La Laguna, Tenerife, Spain\label{iull} \and
        Instituto de Astrof\'isica de Canarias (IAC), E-38200 La Laguna, Tenerife, Spain\label{iiac} \and
        Department of Physical Sciences, Ritsumeikan University, Kusatsu, Shiga 525-8577, Japan \label{iur} \and
        Komaba Institute for Science, The University of Tokyo, 3-8-1 Komaba, Meguro, Tokyo 153-8902, Japan
        \label{ikis} \and
        Department of Multi-Disciplinary Sciences, Graduate School of Arts and Sciences, The University of Tokyo, 3-8-1 Komaba, Meguro, Tokyo 153-8902, Japan\label{iut} \and
        Okayama Observatory, Kyoto University, 3037-5 Honjo, Kamogatacho, Asakuchi, Okayama 719-0232, Japan \label{ioo} \and
        Astrobiology Center, 2-21-1 Osawa, Mitaka, Tokyo 181-8588, Japan \label{ica}
        \and
        National Astronomical Observatory of Japan, 2-21-1 Osawa, Mitaka, Tokyo 181-8588, Japan\label{naoj}
        }
            
   \date{}
 
  \abstract
   {Rocky planets on ultra-short period orbits can have surface magma oceans and rock-vapour atmospheres in which dust can condense.  Observations of that dust can inform about the composition surface conditions on these objects.} {We constrain the properties and long-term (decade) behaviour of the transiting dust cloud from the ``evaporating" planet K2-22b.} {We observed K2-22b around 40 predicted transits with MuSCAT ground-based multi-optical channel imagers, and  complemented these data with long-term monitoring by the ground-based ATLAS (2018-2024) and space-based \tess{} (2021-2023) surveys.} {We detected signals during 7 transits, none of which showed significant wavelength dependence.  The expected number of MuSCAT-detected transits is $\ge22$, indicating a decline in mean transit depth since the \ktwo{} discovery observations in 2014.}{Lack of significant wavelength dependence indicates that dust grains are large or the cloud is optically thick.  Long-term trends of depth could be due to a magnetic cycle on the host star or overturn of the planet's dayside surface magma ocean.  The possibility that K2-22b is disappearing altogether is ruled out by the stability of the transit ephemeris against non-gravitational forces, which constrains the mass to be at least comparable to Ceres.}
   \maketitle
   
\section{Introduction}

On planets with ultra-short period ($\lesssim$1 d) orbits (USPs) and equilibrium temperatures $>2000$~K, the constitutive elements of silicate minerals (e.g., Si, Mg, Fe) are volatile and can contribute to an atmosphere \citep{Fegley2023}.  Rocky planets that are depleted of highly volatile light elements (H, He, C, O, N) due to their proximity to the host star could instead have rock-vapour atmospheres in equilibrium with magma oceans \citep{Kite2016,Chao2021}.  Escape of these atmospheres over Gyr lowers planet mass and surface gravity, and can lead to catastrophic runaway ``evaporation" of the entire planet \citep{Perez-Becker2013,Curry2024}.

Scattering by dust condensing in the escaping rock-vapour atmosphere of such objects has been invoked to explain rare quasi-periodic transit-like signals discovered in \kepler{} and \ktwo{} time-series photometry of three main sequence stars: KIC 12557548 aka Kepler-1520, KOI-2700, and K2-22 \citep{Rappaport2012,Rappaport2014a,Sanchis-Ojeda2015}.   While these objects dim at strictly periodic intervals, the events vary significantly and stochastically in depth and are sometimes absent.  The duration and shape of the lightcurves also deviates from that of a transiting planet, suggesting an extended dust cloud with a trailing (and sometimes leading) tail \citep{vanLieshout2016}.   The progenitor planets are presumably transiting but the absence of a detectable signal at some epochs indicates they are smaller than Mercury \citep{Perez-Becker2013}.  These lightcurves are analogous to those of some ``dipper" stars, but the latter phenomenon is more pronounced and is invariably associated with T Tauri disks \citep[e.g.,][]{Ansdell2016a}, with rare exceptions \citep{Gaidos2019b,Gaidos2022b}.  

Observations of evaporating rocky planets can probe their otherwise inaccessible interior composition  \citep{Bodman2018,Okuya2020,Zilinskas2022} and test theories of planet formation and migration under extreme conditions close to the star \citep{Winn2018,Adams2021}.  Monitoring of variation in transit depth can test models of chaotic evaporation and dust production feedbacks \citep{Bromley2023,Booth2023}.  Since Mie-like scattering by dust is wavelength-dependent, multi-band photometry can constrain dust grain size \citep{Croll2014,Bochinski2015,Sanchis-Ojeda2015,Colon2018,Ridden-Harper2019,Schlawin2021}, a key parameter of outflow models used for lightcurve and mass loss calculations \citep{Perez-Becker2013,Campos-Estrada2024}.  

We describe long-term, multi-band optical monitoring from ground and space of the evaporating planet K2-22b, discovered in data obtained by \ktwo\ during Campaign 1 in 2014 \citep{Sanchis-Ojeda2015}.   The (unseen) object is on a 9.14 hr-orbit around a K7-type main sequence dwarf star.  Using ground-based monitoring, \citet{Colon2018} found that transits persisted for at least several years after the discovery.   \citet{Sanchis-Ojeda2015}, \citet{Colon2018}, and \citet{Schlawin2021} found no wavelength dependence in the depth and duration of most events, but a greater depth at shorter wavelengths for the deepest transits.  This resembles the lack of wavelength dependence seen among the transits of Kepler-1520b \citep{Croll2014}.  Spectroscopic searches for accompanying gas, i.e. volatilised neutral sodium, have not yielded detections \citep{Gaidos2019a,Ridden-Harper2019}.   

\section{Observations and Data Reduction}
\label{sec:observations}

Between December 2021 and April 2023 we observed 33 predicted transits of K2-22b with the \mthree imager \citep{Narita2020} installed at the 2\,m Faulkes Telescope North at Haleakala Observatory on Maui, Hawai'i operated by the Las Cumbres Observatory Global Telescope (LCOGT), and 7 transits with the \mtwo imager \citep{Narita2019} on the 1.5\,m Telescopio Carlos Sánchez at the Teide Observatory, Spain (Table \ref{tab:observations}).  Both \mtwo and \mthree observe simultaneously and either synchronously (same cadence) or asynchronously in $gri$ and $z_s$ (short) bands.  The star was typically observed for 3 hr centred on the predicted 46-min long transit with a cadence between 0.5 and 2 min, depending on mode and filter.  The window for 16 observations was 1.5 hr early due to an error in the ephemeris, nevertheless most of these included whole or partial transit intervals.  Nine observations were rendered useless due to weather and a tenth due to guiding problems, and two others missed the transit completely.  \mtwo data were analysed using a dedicated pipeline that performs image processing and standard aperture photometry and calibration as described in \citet{Parviainen2020}, while the \mthree observations were first processed using the LCOGT BANZAI pipeline \citep{McCully2018}, after which photometry was extracted using the \mtwo pipeline.  

\begin{table}
\centering
\caption{MuSCAT Observations of K2-22}
\begin{tabular*}{\columnwidth}{@{\extracolsep{\fill}} llll}
\toprule
\toprule
\multicolumn{1}{c}{UT Date (start)} & \multicolumn{1}{c}{images} & \multicolumn{1}{c}{depth} & Note\\
& \multicolumn{1}{c}{($g'/r'/i'/Z_s$)} & \multicolumn{1}{c}{(ppm)$^c$} & \\
\midrule
\midrule
\multicolumn{4}{c}{M3 ($g/r/i/z$ exposure times all 120 sec)}\\
\midrule
2020-12-29 10:31$^{a}$ & 83 & --- & weather \\
2021-01-06 10:04$^{a}$ & 85 & --- & guiding \\
2021-01-09 11:13$^{a}$ & 82 & --- & weather \\
2021-01-11 08:57$^{a}$ & 70 & --- & weather \\
2021-01-22 10:09$^{a}$ & 85 & --- & weather \\
2021-03-31 06:32$^{a}$ & 83 & $<580$ & \\
2021-04-03 07:42$^{a}$ & 83 & --- & weather \\
2021-04-06 08:52$^{a}$ & 83 & $<370$ & \\
2021-04-16 06:40$^{a}$ & 83 & --- & weather \\
2021-04-19 07:50$^{a}$ & 83 & $<420$ & \\
2021-04-29 05:38$^{a}$ & 82 & --- & missed\\
2021-05-10 06:53$^{a}$ & 83 & $<620$ & \\
2021-05-15 05:47$^{a}$ & 83 & $<730$ & \\
2021-05-18 06:57$^{a}$ & 82 & $<840$ \\
2021-05-23 05:52$^{a}$ & 84 & $<780$ \\
2021-05-26 07:02$^{a}$ & 65 & --- & missed\\
2022-02-08 07:52 & 63 & 3670--4470 & \\
2022-02-27 09:08 & 83 & $<1000$ & \\
2022-03-12 08:06 & 83 & $<710$ & \\
\midrule
\multicolumn{4}{c}{M3 ($g/r/i/z$ exposure times = 122/32/29/55 sec)}\\
\midrule
2022-02-11 09:02 & 83/308/283/181 & 2220--2960 & \\
2022-02-24 07:59 & 83/308/282/181 & 3560--4270 & \\
\midrule
\multicolumn{4}{c}{M3 ($g/r/i/z$ exposure times = 120/30/25/35 sec)}\\
\midrule
2022-03-01 06:52 & 74/74/74/74    & $<460$ & \\
2022-03-05 11:28 & 84/297/346/275 & $<600$ & \\
2022-03-07 09:13 & 83/293/340/270 & --- & weather \\
2022-03-23 09:20 & 84/297/345/275 & $<700$ & \\
2022-03-25 07:04 & 84/297/346/275 & $<820$ & \\
2022-03-28 08:14 & 84/297/346/275 & 1720--2470 & \\
2023-01-21 11:52 & 82/293/340/270 & $<820$ &\\
2023-03-17 08:52 & 84/295/342/274 & --- & weather \\
2023-03-27 06:39 & 84/203/339/271 & $<340$ & \\
2023-04-07 07:53 & 83/284/330/263 & $<1580$ & \\
2023-04-10 09:03 & 84/298/348/276 & 1390--2170 &  \\
2023-04-23 08:00 & 84/299/348/277 & $<270$ &\\
\midrule
\multicolumn{4}{c}{M2 ($g/r/i/z$ exposure times = 15/90/90/90 sec)}\\
\midrule
2021-01-22 01:03 & 1188/313/211/211 & 1300--1510 & \\ 
2021-02-12 00:23 & 697/124/124/124 & $<$800 &  \\ 
2021-12-18 03:10 & 782/137/-$^{b}$/142 & $<$990 & \\ 
2021-12-23 02:30 & 477/48/-$^{b}$/91 & $<$1660 & \\ 
2021-12-24 02:37 & 269/15/-$^{b}$/154 & --- & weather\\ 
2022-01-06 02:20 & 917/163/163/163 & 1680--2840 & \\ 
2022-01-08 01:59 & 685/140/131/139 & $<$3800 & \\ 
\bottomrule
\end{tabular*}
\tablefoot{\tablefoottext{a}{Ephemeris error of 1.5 hr meant that transits occurred at the end of each interval.} \tablefoottext{b}{The M2 $i$-band camera was out of order in December 2021.} \tablefoottext{c}{68\% confidence interval or upper limit on achromatic transit depth.}}
\label{tab:observations}
\end{table}

\section{Analysis and Numerical Methods}
\label{sec:analysis}

To set limits on individual transit depths $\Delta F$, we fit a transit lightcurve model constructed with \pytransit \citep{Parviainen2015,Parviainen2020a, Parviainen2020b} to our MuSCAT photometry. The model used different limb-darkening parameters for each bandpass, but a wavelength-independent transit depth. The Bayesian fit adopted normal priors based on the values of \citet{Sanchis-Ojeda2015} for the orbital parameters (time of inferior conjunction, orbital period, inclination, and stellar density), and \ldtk \citep{Parviainen2015b} with the stellar parameters of \citet{Sanchis-Ojeda2015} for the two limb-darkening coefficients in each passband.

Observations with a significant transit detection were jointly analysed by a model in which orbital and limb-darkening parameters are the same for all transits, but transit depth at a reference wavelength and wavelength dependence are allowed to vary.  The transit depth and effective planet-star area ratio $k^2$ for transit $i$ and passband $j$ is defined as
\begin{equation}
    \Delta F_\mathrm{ij} \approx k^2_\mathrm{ij} = \frac{\int T_\mathrm{j}(\lambda)\; k^2_\mathrm{i0}\; e^{-\alpha_\mathrm{i} (\lambda - \lambda_0)}\; \ud\lambda}{\int T_\mathrm{j}(\lambda)\;\ud\lambda},
\end{equation}
where $T$ is the bandpass response function, $k^2_\mathrm{i0}$ is the star-planet area ratio at a reference wavelength $\lambda_0$, and $\alpha$ is the \Angstrom exponent describing the wavelength dependence of $k^2$.  
  
\section{Results}
\label{sec:results}

Five \mthree observations and two \mtwo observations revealed a significant transit-like signal around the predicted times (Table \ref{tab:observations} and Fig. \ref{fig:all}).   We combined these detections with previous measurements \citep{Sanchis-Ojeda2015,Colon2018,Schlawin2021} to revise the ephemeris: $T_c = 2456811.1207 \pm 0.0006$ (BJD) and $P = 0.38107710 \pm 1\times10^{-7}$~d.

\begin{figure*}
    \centering
    \includegraphics[width=\linewidth]{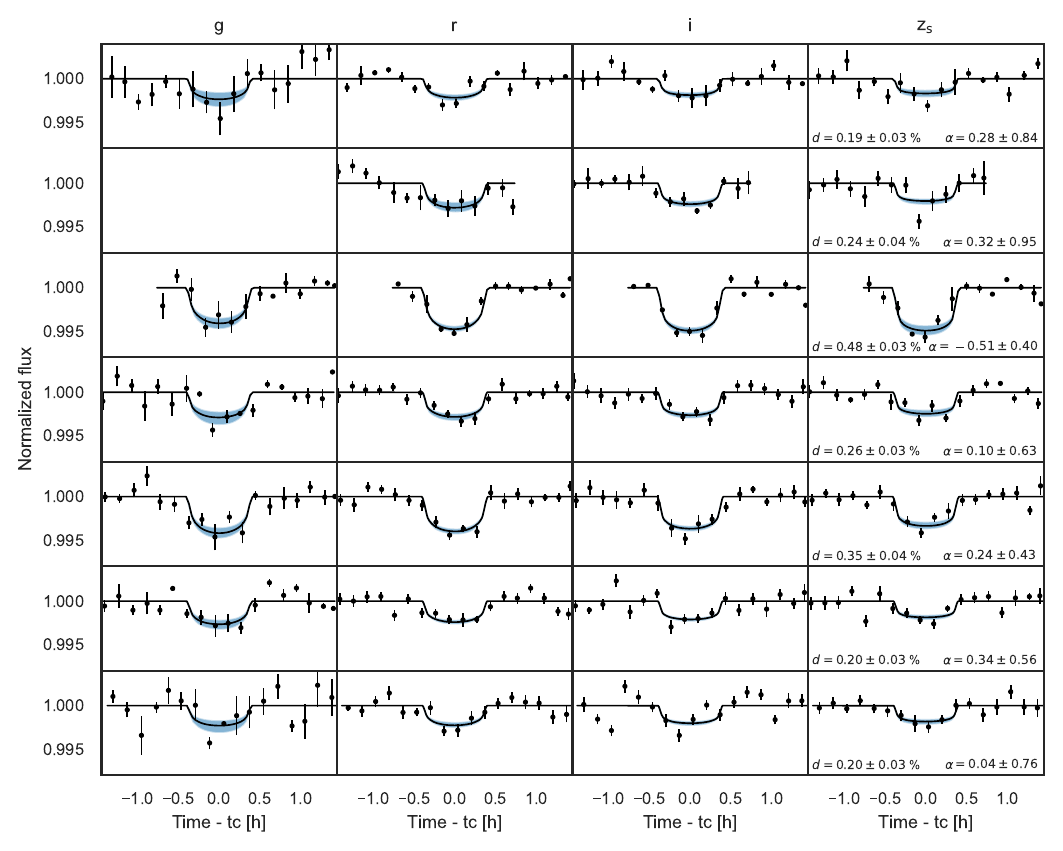}
    \caption{\mthree and \mtwo photometry of seven K2-22b transits (binned to 10 min intervals for visualisation, black points) is plotted with the median of the posterior model fits (black line) and 16th and 84th percentiles (blue shading),  The posteriors of the reference band transit depth $k_0 = R_p/R_*$ and \Angstrom coefficient $\alpha$ estimate are reported in the rightmost panels.}
    \label{fig:fit}
\end{figure*}

\begin{figure}
    \centering
    \includegraphics[width=\columnwidth]{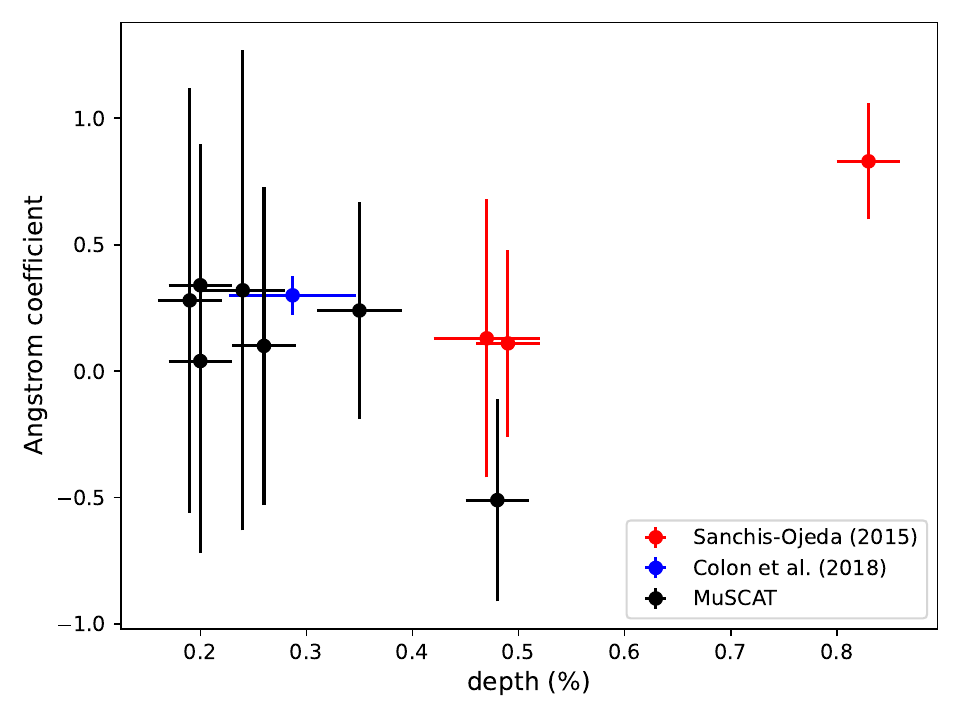}
    \caption{\Angstrom coefficients of transits of K2-22b from MuSCAT photometry plus previously published values.  The value from \citet{Colon2018} was determined from ``blue" and ``red" synthetic photometry (their Fig. 7a) derived from GTC-OSIRIS time-series spectra.}
    \label{fig:angstrom}
\end{figure}

\begin{figure*}[ht]
    \centering
    \includegraphics[width=\textwidth]{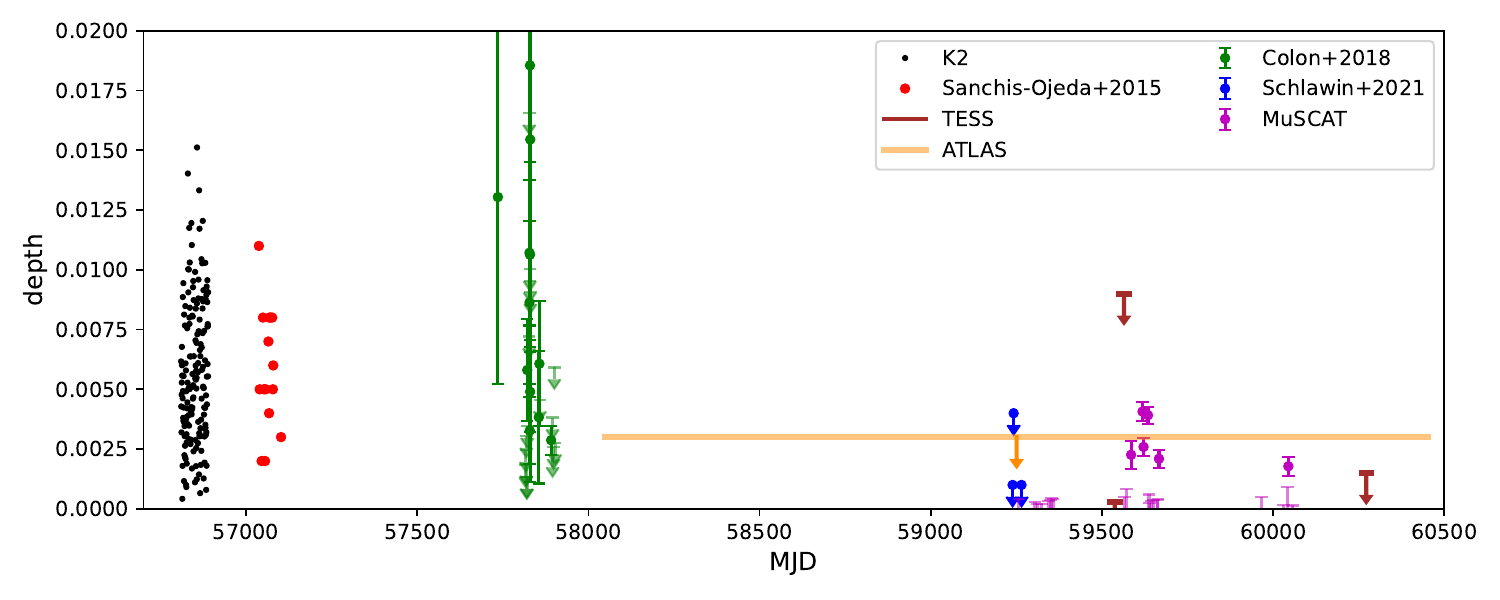}
    \caption{Compilation of depths of detections and upper limits of transits of K2-22b.  \ktwo{} values are as re-derived in this work and \citet{Sanchis-Ojeda2015} values are from ground-based follow-up.  The \tess{} and ATLAS values are $p=0.01$ upper limits derived from an entire sector/survey (see Fig. \ref{fig:all} for ATLAS lightcurve).}
    \label{fig:all}
\end{figure*}

We retrieved values of \Angstrom coefficient $\alpha$ for each detected transit.  Figure \ref{fig:angstrom} plots these with values for three transits in 2015 from \citet{Sanchis-Ojeda2015}.  We also estimated an $\alpha$ of $0.30 \pm 0.08$ from the synthetic photometry that \citet{Colon2018} constructed from GTC-OSIRIS time-series spectroscopy (top panel of their Fig. 7).  To estimate transit depth, we averaged values around the deepest part of the transit and used the mean wavelengths in the blue and red OSIRIS intervals (623 and 808 nm, respectively).  We estimated the uncertainty using 1000 Monte Carlo simulations of the data, adopting the standard deviations in the transit intervals as the errors.  Only the original deep transit analysed by \citet{Sanchis-Ojeda2015} has a significantly positive value of $\alpha$.  When compared to models of single Mie scattering, these values of $\alpha$ suggest grains larger than 0.3 \micron{} (Appendix \ref{app:scattering}).         

Lightcurves of K2-22 from Sectors 45, 46, and 72 of the Transiting Exoplanet Survey Satellite \citep[\tess,][]{Ricker2015} covering 80 days (Fig. \ref{fig:timeline} in Appendix \ref{sec:tessatlas}), phased to the ephemeris of ``b", exhibit no significant indication of transits (Fig. \ref{fig:tess}).  We established upper limits ($p=0.01$ significance) of 0.03, 0.09, and 0.15\% for the respective sectors (Fig. \ref{fig:all} and Appendix \ref{sec:tessatlas}).  Likewise, 6 years of photometry from the Asteroid Terrestrial-impact Last Alert System \citep[ATLAS,][]{Tonry2018} phased to the period of K2-22b (Fig. \ref{fig:atlas}) contain no significant signal at inferior conjunction (Appendix \ref{sec:tessatlas}).  The upper limits are less restrictive than \tess{} (2\% and 0.3\% in $c$- and $o$-bands, respectively) but the baseline is much longer (Fig. \ref{fig:all} and Appendix \ref{sec:tessatlas}).

The absence of a detectable average signal in \tess{} or ATLAS is not due to ephemeris error since over 6.5 yr, the period error of $1 \times 10^{-7}$~d amounts to an ephemeris error of only 1 min or 0.004 in phase.   Since the \tess{} (but not the ATLAS) passband is somewhat redder than \kepler{}, scattering by a dust cloud should be weaker.  However,  calculations of this effect using the constraint on grain size from our MuSCAT photometry limit this to $\lesssim$10\% (Appendix \ref{app:scattering}).  Our new ephemeris agrees with that of \citet{Schlawin2021} and is more accurate than but consistent with \citet{Sanchis-Ojeda2015}.  The use of the latter would not appreciably change the \tess{} or ATLAS phased lightcurves.

\begin{figure}[ht]
    \centering
    \includegraphics[width=\columnwidth]{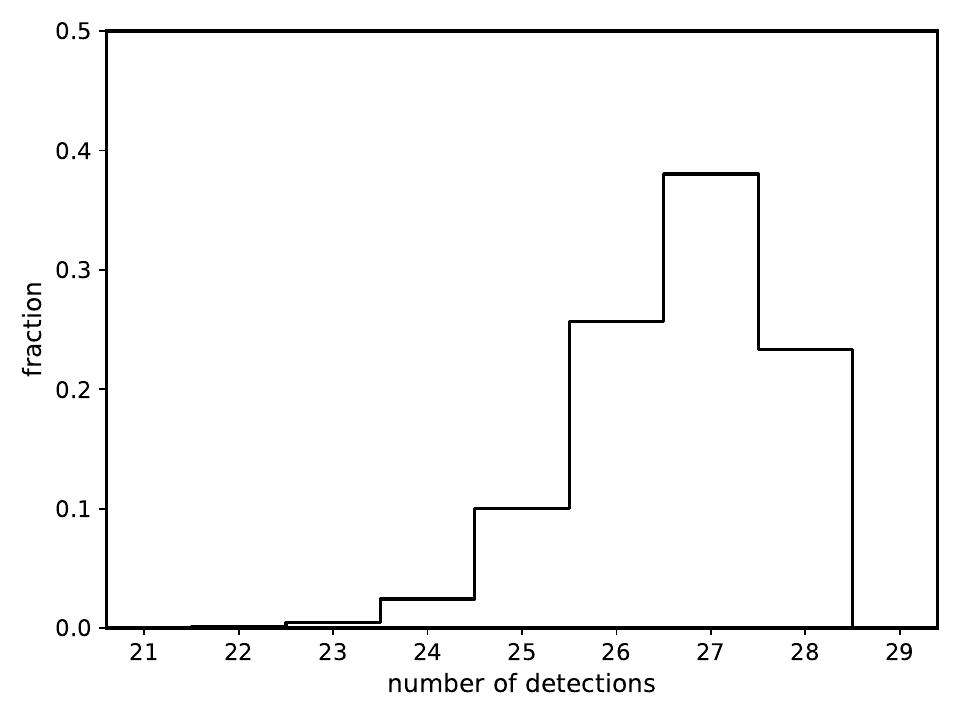}
    \caption{Fractional distribution of predicted number of detections with MuSCAT (of 28 possible) in $10^4$ Monte Carlo simulations based on the distribution of transit depths observed by \ktwo{} in 2014.  The actual number of MuSCAT detections is seven.}
    \label{fig:montecarlo}
\end{figure}

To statistically assess the decline in transits of K2-22b, we computed the expected number of MuSCAT detections based on the distribution of transit depths in the \ktwo{} discovery lightcurve.  Because \citet{Sanchis-Ojeda2015} reported a subset of events with satisfactory lightcurve fits we performed a fit of the entire \ktwo{} PDC-SAP lightcurve using {\tt PyTransit} \citep{Parviainen2015}.  The lightcurve was detrended using a linearly-interpolated median in a 25-point (12.5-hour) window, where the upper and lower 16\% of the points in each bin are excluded.  Transit central times were computed using our revised ephemeris, and model lightcurves were computed using limb darkening parameters from the {\tt LDTk} toolkit \citep{Parviainen2015b} based on the stellar parameters of \citet{Sanchis-Ojeda2015}.  These calculations adopted a circular orbit with $a/R_* = 3.3$ \citep{Sanchis-Ojeda2015} and accounted for the 30-min integration time of \ktwo.  These models were fit to the data varying only $R_p/R_*$ and impact parameter (i.e. orbital inclination).  Using the best-fit parameter values, the transit depth was re-evaluated for the shorter integration times of the MuSCAT observations (few minutes).  In the minority of cases where the fit failed (inclinations $>$90 deg or less than the value for unit impact parameter, or transit depths $>0.02$), we set the transit depth to one minus the minimum in the lightcurve immediately around the predicted central time.  The depth distribution of 171 transits resembles that of \citet{Sanchis-Ojeda2015} and has a mean of 0.56\% (Fig. \ref{fig:all}).  We performed $10^4$ Monte-Carlo simulations of our MuSCAT campaign, with each observation in the campaign assigned a transit depth drawn from an interpolated \ktwo{} distribution.  We flagged as ``detections" those with depths that exceed either the upper limit (for actual non-detections) or the lower bound (for actual detections).  Figure \ref{fig:montecarlo} shows the distribution of the number of simulated detections in our Monte-Carlo campaigns.  Compared to 28 possible detections, the predicted minimum, median and mode are 22, 26.5, and 27.  The actual number of 7, and this distribution rules out the possibility that the intrinsic depth distribution is unchanged and we had few detections by chance.  Also, the upper limits from the three \tess{} sectors and ATLAS $o$-band photometry are all below the mean \ktwo{} transit depth.  

\section{Conclusions and Discussion}

\subsection{Dust properties}

The absence of significant wavelength-dependence ($\alpha > 0$) in our transit depth measurements agrees with previous findings \citep{Sanchis-Ojeda2015,Colon2015,Schlawin2021}. When compared to single Mie scattering models, our MuSCAT-based \Angstrom coefficients suggest grain sizes $>0.3$ \micron{} (see Appendix).  Alternatively, multiple scattering is occurring in an optically thick cloud.  The existence of a leading dust tail in the orbit of K2-22b also indicates that radiation pressure is small compared to the gravitational force, implying either very small or very large dust grains \citep{Sanchis-Ojeda2015}. \citet{Campos-Estrada2024} predicts that that the K2-22b dust cloud has an optical depth $\tau \sim$1-10 along lines of sight near the planet, and thus appreciable \emph{vertical} optical depth.  The latter -- and commensurate surface cooling -- is invoked to explain the variability between transit events as a limit cycle between dust formation and magma pool evaporation \citep{Booth2023,Bromley2023}.  A \micron{} size for dust raises the question of how such grains are lofted in the expected vapour wind \citep{Perez-Becker2013}, and challenges a model of chaotic behaviour that assumes very small dust \citep{Bromley2023}.

\subsection{Intermittency of the K2-22b Transit Signal}

A change in the dust cloud over the past decade, distinct from the inter-transit variability modelled by \citet{Booth2023,Bromley2023}, is analogous to the years-long variation in the signal of Kepler-1520b reported by \citet{Rappaport2012,Schlawin2018}.   We consider two scenarios to explain this change: variability of the host star, or intrinsic changes on the planet.  

One mechanism involving the star involves the effect of its magnetic activity and wind on the dust tail from the planet.  \citet{Kawahara2013} found an anti-correlation between the transit depth of Kepler-1520b and stellar rotational variability (derived from the same \kepler{} lightcurve).  This could be evidence for a link, although the effect of spots on apparent transit depth may play a role \citep{Croll2015,Schlawin2018}.  For \micron-size dust, the stellar wind pressure is predicted to be much less than the radiation pressure, but if grains are charged, Lorenz forces could be important \citep{Price2019,Price2023}.  The majority of M dwarfs (K2-22 has a K7 spectral type) exhibit cycle-like variability in magnetic activity \citep{Mignon2023}.  Where magnetic cycles on cool dwarfs are detected,  their periods range a couple of years to over a decade, with a positive correlation with rotation period \citep{Saar1999,Suarez-Mascareno2016} but no discernible trend with spectral type \citep{Suarez-Mascareno2016,Mignon2023}.  

As has been suggested for Kepler-1520b \citep{Kite2016}, long-term change in the behaviour of K2-22b could reflect variability in the surface composition of the substellar magma pool thought to be the source of the dust-forming wind.  \citet{Kite2016} described four regimes for magma pool coupled to such a wind depending on the stellar irradiance and the Fe content of the silicate mantle.  On relatively Fe-poor (including Earth-like) planets, vaporisation of the magma pool produces a negatively-buoyant chemical boundary layer or ``lag".  At the same time, the thermal gradient between the substellar point and the pool's edge is expected to drive a global-scale overturning flow.  If the timescale for lag formation is much shorter than the overturn timescale then the magma pool will be ``patchy" and the observable average of many such patches would change little.  If the reverse is true, the magma pool will be chemically uniform.  However, if the timescales are comparable then the surface of the magma pool could chemically vary on the overturning timescale.  This variation could be amplified by the sensitivity of dust grain opacity to Fe content and equilibrium vapour pressure to temperature \citep{Bromley2023}.  \citet{Kite2016} relate the overturning circulation to thermal buoyancy forces and find a timescale of order a decade (their Eqn. 5),  consistent with the timescale of observed change for K2-22b.  If indeed the magma pool is circulating, this means that its depth is very shallow (a multiple of the thermal boundary layer) and its chemistry is not representative of the bulk crust or mantle \cite{Kite2016,Curry2024}.

A third explanation is that the source of dust is not a planet but a much smaller body which is in terminal decline.  The stability of the transit phase over a decade ($\sim10^4$ orbits) unambiguously rules out scenarios where dust is temporarily trapped in the stellar magnetic field \citep{Sanderson2023,Bouma2024}.  Moreover, it limits drift in phase due to non-gravitational acceleration to $<5 \times 10^{-3}$, corresponding to $4 \times 10^4$ km, over a decade.  This places an upper limit on the acceleration (of $<10^{-12}$ km sec$^{-2}$) and a lower limit on its mass.  Such logic has been used to estimate the masses of comet nuclei \citep[][ and references therein]{Sosa2009}.  Adopting a mass loss rate of $\sim2$ \mearth{} Gyr$^{-1}$ \citep{Schlawin2021,Campos-Estrada2024}, and a net velocity (averaged over the surface) equal to $\xi v_{\rm th}$ where $v_{\rm th}$ is the thermal speed ($\sim$2 km s$^{-1}$) and $\xi$ is a dimensionless factor that is $\sim$0.5 for comets \citep{Sosa2009} -- but here conservatively taken to be 0.1 -- the mass of K2-22b must be $>1.5\times 10^{-5}$ \mearth{} or 10\% the mass of Ceres.   At the current mass loss rates such a body would not evaporate for $\gtrsim10^4$~yr and it is extremely unlikely we would be  observing it right at its demise.  

More long-term monitoring of K2-22b is needed to definitively test these scenarios.  If the variation is due to stellar activity or overturn of a surface magma ocean, transits of K2-22b will eventually become more frequent.  If it is the continuous disintegration of a much smaller dust source, dimming events will disappear and not return.   Since \tess{} will not re-observe K2-22 in the foreseeable future, nor is it in either of the \emph{PLATO} long-stare fields \citep{Nascimbeni2022}, monitoring from the ground, e.g., by ATLAS is key.   Any correlation with the magnetic activity of the star could be elucidated by spectroscopy of chromospheric emission lines.

If objects such as K2-22b only produce dust clouds with a finite duty cycle, then their intrinsic occurrence is higher than previously estimated, aggravating a possible discrepancy between the occurrence of evaporating planets and the USP population presumed to be the progenitors \citep{Curry2024}.  Long-term variation in the behaviour of K2-22b (and other evaporating planets) could also impact planning for follow-up campaigns with major observatories.

\begin{acknowledgements} 
EG was supported by NASA Astrophysics Data Analysis award 80NSSC19K0587 and US National Science Foundation Astronomy \& Astrophysics Grant 2106927, and was a visiting scientist at the International Space Sciences Institute in Bern during a portion of this work.  HP acknowledges support from the Spanish Ministry of Science and Innovation with the Ramon y Cajal fellowship number RYC2021-031798-I, and funding from the University of La Laguna and the Spanish Ministry of Universities. This work is partly supported by JSPS KAKENHI Grant Numbers JP21K13955, JP24H00017, JP24K00689, JP24K17083 and JSPS Bilateral Program Number JPJSBP120249910. This article is partly based on observations made with the MuSCAT2 and MuSCAT3 instruments, developed by the AstroBiology Center (ABC) under the National Institutes of Natural Sciences of Japan with support by ABC, JSPS KAKENHI (JP18H05439), and JST PRESTO (JPMJPR1775).  MuSCAT2 is operated by the IAC in the Spanish Observatorio del Teide, while MuSCAT3 on the Faulkes Telescope North, is operated by the LCOGT network Part of the	LCOGT telescope time was granted by NOIRLab through the Mid-Scale Innovations Program (MSIP). MSIP is funded by NSF. This work has made use of data from the Asteroid Terrestrial-impact Last Alert System (ATLAS) project. The ATLAS project is primarily funded to search for near earth asteroids through NASA grants NN12AR55G, 80NSSC18K0284, and 80NSSC18K1575; byproducts of the NEO search include images and catalogues from the survey area.  This paper includes data collected by the \tess{} mission, funding of which is provided by the NASA's Science Mission Directorate.  This work used the {\tt SciPy} and {\tt AstroPy} packages \citep{SciPy2020,astropy2022}.  

\end{acknowledgements}


\begin{thebibliography}{57}
\expandafter\ifx\csname natexlab\endcsname\relax\def\natexlab#1{#1}\fi

\bibitem[{{Adams} {et~al.}(2021){Adams}, {Jackson}, {Johnson}, {Ciardi},
  {Cochran}, {Endl}, {Everett}, {Furlan}, {Howell}, {Jayanthi}, {MacQueen},
  {Matson}, {Partyka-Worley}, {Schlieder}, {Scott}, {Stanton}, \&
  {Ziegler}}]{Adams2021}
{Adams}, E.~R., {Jackson}, B., {Johnson}, S., {et~al.} 2021, \psj, 2, 152

\bibitem[{{Ag{\"u}eros} {et~al.}(2018){Ag{\"u}eros}, {Bowsher}, {Bochanski},
  {Cargile}, {Covey}, {Douglas}, {Kraus}, {Kundert}, {Law}, {Ahmadi}, \&
  {Arce}}]{Agueros2018}
{Ag{\"u}eros}, M.~A., {Bowsher}, E.~C., {Bochanski}, J.~J., {et~al.} 2018,
  \apj, 862, 33

\bibitem[{{Ansdell} {et~al.}(2016){Ansdell}, {Gaidos}, {Rappaport}, {Jacobs},
  {LaCourse}, {Jek}, {Mann}, {Wyatt}, {Kennedy}, {Williams}, \&
  {Boyajian}}]{Ansdell2016a}
{Ansdell}, M., {Gaidos}, E., {Rappaport}, S.~A., {et~al.} 2016, \apj, 816, 69

\bibitem[{{Astropy Collaboration} {et~al.}(2022){Astropy Collaboration},
  {Price-Whelan}, {Lim}, {Earl}, {Starkman}, {Bradley}, {Shupe}, {Patil},
  {Corrales}, {Brasseur}, {N{\"o}the}, {Donath}, {Tollerud}, {Morris},
  {Ginsburg}, {Vaher}, {Weaver}, {Tocknell}, {Jamieson}, {van Kerkwijk},
  {Robitaille}, {Merry}, {Bachetti}, {G{\"u}nther}, {Aldcroft},
  {Alvarado-Montes}, {Archibald}, {B{\'o}di}, {Bapat}, {Barentsen},
  {Baz{\'a}n}, {Biswas}, {Boquien}, {Burke}, {Cara}, {Cara}, {Conroy},
  {Conseil}, {Craig}, {Cross}, {Cruz}, {D'Eugenio}, {Dencheva}, {Devillepoix},
  {Dietrich}, {Eigenbrot}, {Erben}, {Ferreira}, {Foreman-Mackey}, {Fox},
  {Freij}, {Garg}, {Geda}, {Glattly}, {Gondhalekar}, {Gordon}, {Grant},
  {Greenfield}, {Groener}, {Guest}, {Gurovich}, {Handberg}, {Hart},
  {Hatfield-Dodds}, {Homeier}, {Hosseinzadeh}, {Jenness}, {Jones}, {Joseph},
  {Kalmbach}, {Karamehmetoglu}, {Ka{\l}uszy{\'n}ski}, {Kelley}, {Kern},
  {Kerzendorf}, {Koch}, {Kulumani}, {Lee}, {Ly}, {Ma}, {MacBride}, {Maljaars},
  {Muna}, {Murphy}, {Norman}, {O'Steen}, {Oman}, {Pacifici}, {Pascual},
  {Pascual-Granado}, {Patil}, {Perren}, {Pickering}, {Rastogi}, {Roulston},
  {Ryan}, {Rykoff}, {Sabater}, {Sakurikar}, {Salgado}, {Sanghi}, {Saunders},
  {Savchenko}, {Schwardt}, {Seifert-Eckert}, {Shih}, {Jain}, {Shukla}, {Sick},
  {Simpson}, {Singanamalla}, {Singer}, {Singhal}, {Sinha}, {Sip{\H{o}}cz},
  {Spitler}, {Stansby}, {Streicher}, {{\v{S}}umak}, {Swinbank}, {Taranu},
  {Tewary}, {Tremblay}, {de Val-Borro}, {Van Kooten}, {Vasovi{\'c}}, {Verma},
  {de Miranda Cardoso}, {Williams}, {Wilson}, {Winkel}, {Wood-Vasey}, {Xue},
  {Yoachim}, {Zhang}, {Zonca}, \& {Astropy Project Contributors}}]{astropy2022}
{Astropy Collaboration}, {Price-Whelan}, A.~M., {Lim}, P.~L., {et~al.} 2022,
  \apj, 935, 167

\bibitem[{{Bochinski} {et~al.}(2015){Bochinski}, {Haswell}, {Marsh}, {Dhillon},
  \& {Littlefair}}]{Bochinski2015}
{Bochinski}, J.~J., {Haswell}, C.~A., {Marsh}, T.~R., {Dhillon}, V.~S., \&
  {Littlefair}, S.~P. 2015, \apjl, 800, L21

\bibitem[{{Bodman} {et~al.}(2018){Bodman}, {Wright}, {Desch}, \&
  {Lisse}}]{Bodman2018}
{Bodman}, E. H.~L., {Wright}, J.~T., {Desch}, S.~J., \& {Lisse}, C.~M. 2018,
  \aj, 156, 173

\bibitem[{{Booth} {et~al.}(2023){Booth}, {Owen}, \& {Schulik}}]{Booth2023}
{Booth}, R.~A., {Owen}, J.~E., \& {Schulik}, M. 2023, \mnras, 518, 1761

\bibitem[{{Bouma} {et~al.}(2024){Bouma}, {Jayaraman}, {Rappaport}, {Rebull},
  {Hillenbrand}, {Winn}, {David-Uraz}, \& {Bakos}}]{Bouma2024}
{Bouma}, L.~G., {Jayaraman}, R., {Rappaport}, S., {et~al.} 2024, \aj, 167, 38

\bibitem[{{Bromley} \& {Chiang}(2023)}]{Bromley2023}
{Bromley}, J. \& {Chiang}, E. 2023, \mnras, 521, 5746

\bibitem[{{Budaj} {et~al.}(2015){Budaj}, {Kocifaj}, {Salmeron}, \&
  {Hubeny}}]{Budaj2015}
{Budaj}, J., {Kocifaj}, M., {Salmeron}, R., \& {Hubeny}, I. 2015, \mnras, 454,
  2

\bibitem[{{Campos Estrada} {et~al.}(2024){Campos Estrada}, {Owen}, {Jankovic},
  {Wilson}, \& {Helling}}]{Campos-Estrada2024}
{Campos Estrada}, B., {Owen}, J.~E., {Jankovic}, M.~R., {Wilson}, A., \&
  {Helling}, C. 2024, \mnras, 528, 1249

\bibitem[{{Chao} {et~al.}(2021){Chao}, {deGraffenried}, {Lach}, {Nelson},
  {Truax}, \& {Gaidos}}]{Chao2021}
{Chao}, K.-H., {deGraffenried}, R., {Lach}, M., {et~al.} 2021, Chemie der Erde
  / Geochemistry, 81, 125735

\bibitem[{{Col{\'o}n} {et~al.}(2015){Col{\'o}n}, {Morehead}, \&
  {Ford}}]{Colon2015}
{Col{\'o}n}, K.~D., {Morehead}, R.~C., \& {Ford}, E.~B. 2015, \mnras, 452, 3001

\bibitem[{{Col{\'o}n} {et~al.}(2018){Col{\'o}n}, {Zhou}, {Shporer}, {Collins},
  {Bieryla}, {Espinoza}, {Murgas}, {Pattarakijwanich}, {Awiphan}, {Armstrong},
  {Bailey}, {Barentsen}, {Bayliss}, {Chakpor}, {Cochran}, {Dhillon}, {Horne},
  {Ireland}, {Kedziora-Chudczer}, {Kielkopf}, {Komonjinda}, {Latham}, {Marsh},
  {Mkrtichian}, {Pall{\'e}}, {Ruffolo}, {Sefako}, {Tinney}, {Wannawichian}, \&
  {Yuma}}]{Colon2018}
{Col{\'o}n}, K.~D., {Zhou}, G., {Shporer}, A., {et~al.} 2018, \aj, 156, 227

\bibitem[{{Croll} {et~al.}(2014){Croll}, {Rappaport}, {DeVore}, {Gilliland},
  {Crepp}, {Howard}, {Star}, {Chiang}, {Levine}, {Jenkins}, {Albert}, {Bonomo},
  {Fortney}, \& {Isaacson}}]{Croll2014}
{Croll}, B., {Rappaport}, S., {DeVore}, J., {et~al.} 2014, \apj, 786, 100

\bibitem[{{Croll} {et~al.}(2015){Croll}, {Rappaport}, \& {Levine}}]{Croll2015}
{Croll}, B., {Rappaport}, S., \& {Levine}, A.~M. 2015, \mnras, 449, 1408

\bibitem[{{Curry} {et~al.}(2024){Curry}, {Booth}, {Owen}, \&
  {Mohanty}}]{Curry2024}
{Curry}, A., {Booth}, R., {Owen}, J.~E., \& {Mohanty}, S. 2024, \mnras, 528,
  4314

\bibitem[{{Fegley} {et~al.}(2023){Fegley}, {Lodders}, \&
  {Jacobson}}]{Fegley2023}
{Fegley}, Bruce, J., {Lodders}, K., \& {Jacobson}, N.~S. 2023, Chemie der Erde
  / Geochemistry, 83, 125961

\bibitem[{{Gaidos} {et~al.}(2023){Gaidos}, {Claytor}, {Dungee}, {Ali}, \&
  {Feiden}}]{Gaidos2023b}
{Gaidos}, E., {Claytor}, Z., {Dungee}, R., {Ali}, A., \& {Feiden}, G.~A. 2023,
  \mnras, 520, 5283

\bibitem[{{Gaidos} {et~al.}(2019{\natexlab{a}}){Gaidos}, {Hirano}, \&
  {Ansdell}}]{Gaidos2019a}
{Gaidos}, E., {Hirano}, T., \& {Ansdell}, M. 2019{\natexlab{a}}, \mnras, 485,
  3876

\bibitem[{{Gaidos} {et~al.}(2022){Gaidos}, {Hirano}, {Kraus}, {Kuzuhara},
  {Zhang}, {Lee}, {Salama}, {Berger}, {Grunblatt}, {Ansdell}, {Liu},
  {Harakawa}, {Hodapp}, {Jacobson}, {Konishi}, {Kotani}, {Kudo}, {Kurokawa},
  {Nishikawa}, {Omiya}, {Serizawa}, {Tamura}, {Ueda}, \&
  {Vievard}}]{Gaidos2022b}
{Gaidos}, E., {Hirano}, T., {Kraus}, A.~L., {et~al.} 2022, \mnras, 512, 583

\bibitem[{{Gaidos} {et~al.}(2019{\natexlab{b}}){Gaidos}, {Jacobs}, {LaCourse},
  {Vanderburg}, {Rappaport}, {Berger}, {Pearce}, {Mann}, {Weiss}, {Fulton},
  {Behmard}, {Howard}, {Ansdell}, {Ricker}, {Vanderspek}, {Latham}, {Seager},
  {Winn}, \& {Jenkins}}]{Gaidos2019b}
{Gaidos}, E., {Jacobs}, T., {LaCourse}, D., {et~al.} 2019{\natexlab{b}},
  \mnras, 488, 4465

\bibitem[{{Heinze} {et~al.}(2018){Heinze}, {Tonry}, {Denneau}, {Flewelling},
  {Stalder}, {Rest}, {Smith}, {Smartt}, \& {Weiland}}]{Heinze2018}
{Heinze}, A.~N., {Tonry}, J.~L., {Denneau}, L., {et~al.} 2018, \aj, 156, 241

\bibitem[{{Kawahara} {et~al.}(2013){Kawahara}, {Hirano}, {Kurosaki}, {Ito}, \&
  {Ikoma}}]{Kawahara2013}
{Kawahara}, H., {Hirano}, T., {Kurosaki}, K., {Ito}, Y., \& {Ikoma}, M. 2013,
  \apjl, 776, L6

\bibitem[{{Kesseli} {et~al.}(2016){Kesseli}, {Petkova}, {Wood}, {Whitney},
  {Hillenbrand}, {Gregory}, {Stauffer}, {Morales-Calderon}, {Rebull}, \&
  {Alencar}}]{Kesseli2016}
{Kesseli}, A.~Y., {Petkova}, M.~A., {Wood}, K., {et~al.} 2016, \apj, 828, 42

\bibitem[{{Kite} {et~al.}(2016){Kite}, {Fegley}, {Schaefer}, \&
  {Gaidos}}]{Kite2016}
{Kite}, E.~S., {Fegley}, Jr., B., {Schaefer}, L., \& {Gaidos}, E. 2016, \apj,
  828, 80

\bibitem[{{McCully} {et~al.}(2018){McCully}, {Turner}, {Volgenau}, {Harbeck},
  {Valenti}, {Riba}, {Bachelet}, {Snyder}, {Kurczynski}, {Norbury}, \&
  {Street}}]{McCully2018}
{McCully}, C., {Turner}, M., {Volgenau}, N., {et~al.} 2018, {LCOGT/Banzai:
  Initial Release}

\bibitem[{{Mignon} {et~al.}(2023){Mignon}, {Meunier}, {Delfosse}, {Bonfils},
  {Santos}, {Forveille}, {Gaisn{\'e}}, {Astudillo-Defru}, {Lovis}, \&
  {Udry}}]{Mignon2023}
{Mignon}, L., {Meunier}, N., {Delfosse}, X., {et~al.} 2023, \aap, 675, A168

\bibitem[{{Narita} {et~al.}(2019){Narita}, {Fukui}, {Kusakabe}, {Watanabe},
  {Palle}, {Parviainen}, {Monta{\~n}{\'e}s-Rodr{\'\i}guez}, {Murgas},
  {Monelli}, {Aguiar}, {Perez Prieto}, {Oscoz}, {de Leon}, {Mori}, {Tamura},
  {Yamamuro}, {B{\'e}jar}, {Crouzet}, {Hidalgo}, {Klagyivik}, {Luque}, \&
  {Nishiumi}}]{Narita2019}
{Narita}, N., {Fukui}, A., {Kusakabe}, N., {et~al.} 2019, \jatis, 5, 015001

\bibitem[{Narita {et~al.}(2020)Narita, Fukui, Yamamuro, Harbeck, Bowman,
  Elphick, Nation, Armstrong, Han, Abe, Ikoma, Isogai, Kawauchi, Kurita,
  Kusakabe, De~Leon, Livingston, Mori, Nishiumi, Tamura, Watanabe, Volgenau,
  {Heinrich-Josties}, Foale, Daily, McCully, Kirby, Smith, Haworth, Conway,
  {Storrie-Lombardi}, Rosing, Chatelain, Bachelet, Johnson, \&
  Rabus}]{Narita2020}
Narita, N., Fukui, A., Yamamuro, T., {et~al.} 2020, in Society of
  {{Photo-Optical Instrumentation Engineers}} ({{SPIE}}) {{Conference Series}},
  Vol. 11447, 29

\bibitem[{{Nascimbeni} {et~al.}(2022){Nascimbeni}, {Piotto}, {B{\"o}rner},
  {Montalto}, {Marrese}, {Cabrera}, {Marinoni}, {Aerts}, {Altavilla},
  {Benatti}, {Claudi}, {Deleuil}, {Desidera}, {Fabrizio}, {Gizon}, {Goupil},
  {Granata}, {Heras}, {Magrin}, {Malavolta}, {Mas-Hesse}, {Ortolani}, {Pagano},
  {Pollacco}, {Prisinzano}, {Ragazzoni}, {Ramsay}, {Rauer}, \&
  {Udry}}]{Nascimbeni2022}
{Nascimbeni}, V., {Piotto}, G., {B{\"o}rner}, A., {et~al.} 2022, \aap, 658, A31

\bibitem[{{Okuya} {et~al.}(2020){Okuya}, {Okuzumi}, {Ohno}, \&
  {Hirano}}]{Okuya2020}
{Okuya}, A., {Okuzumi}, S., {Ohno}, K., \& {Hirano}, T. 2020, \apj, 901, 171

\bibitem[{{Parviainen}(2015)}]{Parviainen2015}
{Parviainen}, H. 2015, \mnras, 450, 3233

\bibitem[{Parviainen(2020)}]{Parviainen2020b}
Parviainen, H. 2020, Monthly Notices of the Royal Astronomical Society, 499,
  1633

\bibitem[{Parviainen \& Aigrain(2015)}]{Parviainen2015b}
Parviainen, H. \& Aigrain, S. 2015, Monthly Notices of the Royal Astronomical
  Society, 453, 3821

\bibitem[{Parviainen \& Korth(2020)}]{Parviainen2020a}
Parviainen, H. \& Korth, J. 2020, Monthly Notices of the Royal Astronomical
  Society [\eprint[arxiv]{2009.09965}]

\bibitem[{Parviainen {et~al.}(2020)Parviainen, Palle, {Zapatero-Osorio},
  Montanes~Rodriguez, Murgas, Narita, Hidalgo~Soto, B{\'e}jar, Korth, Monelli,
  Casasayas~Barris, Crouzet, De~Leon, Fukui, Hernandez, Klagyivik, Kusakabe,
  Luque, Mori, Nishiumi, {Prieto-Arranz}, Tamura, Watanabe, Burke, Charbonneau,
  Collins, Collins, Conti, Garcia~Soto, Jenkins, Jenkins, Levine, Li, Rinehart,
  Seager, Tenenbaum, Ting, Vanderspek, Vezie, \& Winn}]{Parviainen2020}
Parviainen, H., Palle, E., {Zapatero-Osorio}, M.~R., {et~al.} 2020, Astronomy
  and Astrophysics, 633, A28

\bibitem[{{Pepper} {et~al.}(2007){Pepper}, {Pogge}, {DePoy}, {Marshall},
  {Stanek}, {Stutz}, {Poindexter}, {Siverd}, {O'Brien}, {Trueblood}, \&
  {Trueblood}}]{Pepper2007}
{Pepper}, J., {Pogge}, R.~W., {DePoy}, D.~L., {et~al.} 2007, \pasp, 119, 923

\bibitem[{{Perez-Becker} \& {Chiang}(2013)}]{Perez-Becker2013}
{Perez-Becker}, D. \& {Chiang}, E. 2013, \mnras, 433, 2294

\bibitem[{{Price} {et~al.}(2023){Price}, {Jones}, {Battams}, \&
  {Owens}}]{Price2023}
{Price}, O., {Jones}, G.~H., {Battams}, K., \& {Owens}, M. 2023, \icarus, 389,
  115218

\bibitem[{{Price} {et~al.}(2019){Price}, {Jones}, {Morrill}, {Owens},
  {Battams}, {Morgan}, {Dr{\"u}ckmuller}, \& {Deiries}}]{Price2019}
{Price}, O., {Jones}, G.~H., {Morrill}, J., {et~al.} 2019, \icarus, 319, 540

\bibitem[{{Rappaport} {et~al.}(2014){Rappaport}, {Barclay}, {DeVore}, {Rowe},
  {Sanchis-Ojeda}, \& {Still}}]{Rappaport2014a}
{Rappaport}, S., {Barclay}, T., {DeVore}, J., {et~al.} 2014, \apj, 784, 40

\bibitem[{{Rappaport} {et~al.}(2012){Rappaport}, {Levine}, {Chiang}, {El
  Mellah}, {Jenkins}, {Kalomeni}, {Kite}, {Kotson}, {Nelson},
  {Rousseau-Nepton}, \& {Tran}}]{Rappaport2012}
{Rappaport}, S., {Levine}, A., {Chiang}, E., {et~al.} 2012, \apj, 752, 1

\bibitem[{{Ricker} {et~al.}(2015){Ricker}, {Winn}, {Vanderspek}, {Latham},
  {Bakos}, {Bean}, {Berta-Thompson}, {Brown}, {Buchhave}, {Butler}, {Butler},
  {Chaplin}, {Charbonneau}, {Christensen-Dalsgaard}, {Clampin}, {Deming},
  {Doty}, {De Lee}, {Dressing}, {Dunham}, {Endl}, {Fressin}, {Ge}, {Henning},
  {Holman}, {Howard}, {Ida}, {Jenkins}, {Jernigan}, {Johnson}, {Kaltenegger},
  {Kawai}, {Kjeldsen}, {Laughlin}, {Levine}, {Lin}, {Lissauer}, {MacQueen},
  {Marcy}, {McCullough}, {Morton}, {Narita}, {Paegert}, {Palle}, {Pepe},
  {Pepper}, {Quirrenbach}, {Rinehart}, {Sasselov}, {Sato}, {Seager},
  {Sozzetti}, {Stassun}, {Sullivan}, {Szentgyorgyi}, {Torres}, {Udry}, \&
  {Villasenor}}]{Ricker2015}
{Ricker}, G.~R., {Winn}, J.~N., {Vanderspek}, R., {et~al.} 2015, Journal of
  Astronomical Telescopes, Instruments, and Systems, 1, 014003

\bibitem[{{Ridden-Harper} {et~al.}(2019){Ridden-Harper}, {Snellen}, {Keller},
  \& {Molli{\`e}re}}]{Ridden-Harper2019}
{Ridden-Harper}, A.~R., {Snellen}, I.~A.~G., {Keller}, C.~U., \&
  {Molli{\`e}re}, P. 2019, \aap, 628, A70

\bibitem[{{Saar} \& {Brandenburg}(1999)}]{Saar1999}
{Saar}, S.~H. \& {Brandenburg}, A. 1999, \apj, 524, 295

\bibitem[{{Sanchis-Ojeda} {et~al.}(2015){Sanchis-Ojeda}, {Rappaport},
  {Pall{\`e}}, {Delrez}, {DeVore}, {Gandolfi}, {Fukui}, {Ribas}, {Stassun},
  {Albrecht}, {Dai}, {Gaidos}, {Gillon}, {Hirano}, {Holman}, {Howard},
  {Isaacson}, {Jehin}, {Kuzuhara}, {Mann}, {Marcy}, {Miles-P{\'a}ez},
  {Monta{\~n}{\'e}s-Rodr{\'{\i}}guez}, {Murgas}, {Narita}, {Nowak}, {Onitsuka},
  {Paegert}, {Van Eylen}, {Winn}, \& {Yu}}]{Sanchis-Ojeda2015}
{Sanchis-Ojeda}, R., {Rappaport}, S., {Pall{\`e}}, E., {et~al.} 2015, \apj,
  812, 112

\bibitem[{{Sanderson} {et~al.}(2023){Sanderson}, {Jardine}, {Collier Cameron},
  {Morin}, \& {Donati}}]{Sanderson2023}
{Sanderson}, H., {Jardine}, M., {Collier Cameron}, A., {Morin}, J., \&
  {Donati}, J.~F. 2023, \mnras, 518, 4734

\bibitem[{{Schlawin} {et~al.}(2018){Schlawin}, {Hirano}, {Kawahara}, {Teske},
  {Green}, {Rackham}, {Fraine}, \& {Bushra}}]{Schlawin2018}
{Schlawin}, E., {Hirano}, T., {Kawahara}, H., {et~al.} 2018, \aj, 156, 281

\bibitem[{{Schlawin} {et~al.}(2021){Schlawin}, {Su}, {Herter}, {Ridden-Harper},
  \& {Apai}}]{Schlawin2021}
{Schlawin}, E., {Su}, K. Y.~L., {Herter}, T., {Ridden-Harper}, A., \& {Apai},
  D. 2021, \aj, 162, 57

\bibitem[{{Sosa} \& {Fern{\'a}ndez}(2009)}]{Sosa2009}
{Sosa}, A. \& {Fern{\'a}ndez}, J.~A. 2009, \mnras, 393, 192

\bibitem[{{Su{\'a}rez Mascare{\~n}o} {et~al.}(2016){Su{\'a}rez Mascare{\~n}o},
  {Rebolo}, \& {Gonz{\'a}lez Hern{\'a}ndez}}]{Suarez-Mascareno2016}
{Su{\'a}rez Mascare{\~n}o}, A., {Rebolo}, R., \& {Gonz{\'a}lez Hern{\'a}ndez},
  J.~I. 2016, \aap, 595, A12

\bibitem[{{Tonry} {et~al.}(2018){Tonry}, {Denneau}, {Heinze}, {Stalder},
  {Smith}, {Smartt}, {Stubbs}, {Weiland}, \& {Rest}}]{Tonry2018}
{Tonry}, J.~L., {Denneau}, L., {Heinze}, A.~N., {et~al.} 2018, \pasp, 130,
  064505

\bibitem[{{van Lieshout} {et~al.}(2016){van Lieshout}, {Min}, {Dominik},
  {Brogi}, {de Graaff}, {Hekker}, {Kama}, {Keller}, {Ridden-Harper}, \& {van
  Werkhoven}}]{vanLieshout2016}
{van Lieshout}, R., {Min}, M., {Dominik}, C., {et~al.} 2016, \aap, 596, A32

\bibitem[{Virtanen {et~al.}(2020)Virtanen, Gommers, Oliphant, Haberland, Reddy,
  Cournapeau, Burovski, Peterson, Weckesser, Bright, {van der Walt}, Brett,
  Wilson, Millman, Mayorov, Nelson, Jones, Kern, Larson, Carey, Polat, Feng,
  Moore, {VanderPlas}, Laxalde, Perktold, Cimrman, Henriksen, Quintero, Harris,
  Archibald, Ribeiro, Pedregosa, {van Mulbregt}, \& {SciPy 1.0
  Contributors}}]{SciPy2020}
Virtanen, P., Gommers, R., Oliphant, T.~E., {et~al.} 2020, Nature Methods, 17,
  261

\bibitem[{{Winn} {et~al.}(2018){Winn}, {Sanchis-Ojeda}, \&
  {Rappaport}}]{Winn2018}
{Winn}, J.~N., {Sanchis-Ojeda}, R., \& {Rappaport}, S. 2018, \nar, 83, 37

\bibitem[{{Zilinskas} {et~al.}(2022){Zilinskas}, {van Buchem}, {Miguel},
  {Louca}, {Lupu}, {Zieba}, \& {van Westrenen}}]{Zilinskas2022}
{Zilinskas}, M., {van Buchem}, C.~P.~A., {Miguel}, Y., {et~al.} 2022, \aap,
  661, A126

\end{thebibliography}

\begin{appendix}

\section{\tess{} and ATLAS photometry of K2-22}
\label{sec:tessatlas}

K2-22 appears as TIC 363445338 in the Input Catalogue \citep{Pepper2007} of the Transiting Exoplanet Survey Satellite \citep[\tess,][]{Ricker2015} and was observed during three 27-day sectors (45, 46, and 72) from 6 Nov to 30 Dec 2021, and from 11 Nov to 7 Dec 2023. Two-minute cadence Pre-search Data Conditioning Simple Aperture Photometry (PDC-SAP) lightcurves generated by the \tess{} SPOC pipeline were retrieved from the Mikulski Archive of Space Telescopes (MAST).  These were generated by the \tess-Science Processing Operations Centre (SPOC) pipeline using 3-4 pixel (i.e. 42" across) apertures centred on the star.

While a Lomb-Scargle periodogram analysis did not return significant signals at the orbital period of K2-22b, in all three there was a significant signal at $\sim$5 days.  The rotation period of the primary star has already been established as $\approx$15 days \citep{Sanchis-Ojeda2015}, and \citet{Gaidos2023b} estimate a rotation-based age of $1.1 \pm 0.2$~Gyr at which mid-to late M-type dwarfs may be still rapidly rotating \citep{Agueros2018}.  Thus  we speculate that the 5-day signal is from the M4 dwarf companion.  While the RMS  of the lightcurves (1.6\%) is larger than the expected transit depth ($\sim$0.5\%),  the RMS of the median of binned, phased lightcurves (red curves in Fig. \ref{fig:atlas}) is $<0.1$\% and does not contain any obvious trend ($\chi^2$ of 30-34 for 19 degrees of freedom).  Based on the statistics of the photometry between predicted ingress and egress (vertical dotted lines in Fig. \ref{fig:tess}), in no sector is a significant ($p=0.01$) mean signal detected and 99\% upper limits for the three sectors are 0.03, 0.09, and 0.15\%.  

Since 2018, K2-22 has also been monitored by the Asteroid Terrestrial-impact Last Alert System \citep[ATLAS,]{Tonry2018,Heinze2018}.  Lightcurves in the survey's broad ($\Delta \lambda >210$ nm) ``cyan" ($c$)  and``orange" ($o$) passbands (respective effective wavelengths of $\sim$520 and 660 nm) were generated by ``forced" photometry on ATLAS images.  After filtering points with error codes or errors exceeding 0.1 mag, there were 593 and 2100 measurements spanning 6.5 years in the respective filters.   

\begin{figure*}
\includegraphics[width=\textwidth]{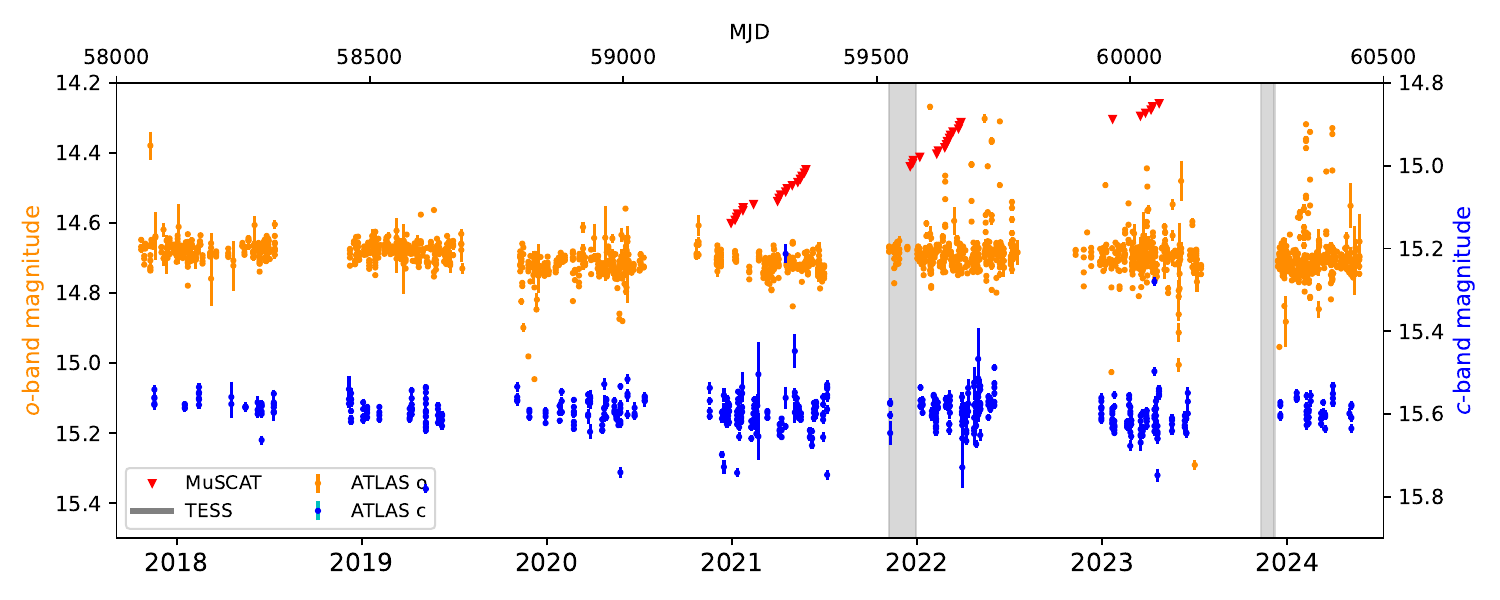}
\caption{ATLAS lightcurves of K2-22 through "$o$" (orange, left y axis) and "$c$" (blue, right y axis) filters.  Vertical grey bars mark observations by \tess{} during Sectors 45-46 and 72.  Inverted red triangles mark MuSCAT observations, sequentially displaced for clarity.  (K2-22 was discovered in mid-2014.)}
\label{fig:timeline}
\end{figure*}

Both lightcurves contain apparent dimming events, including a 0.6-mag drop in $o$-band near the end of the 2023 season (Fig. \ref{fig:all}).  These could be systematics but there is no correlation with either sky brightness or the FHWM of the point spread function for the forced photometry.  The $o$-band lightcurve (but not the $c$-band lightcurve) also contains numerous positive excursions, which we speculate to be due to flares from the mid M-type companion of K2-22 \citep[separation 2",][]{Sanchis-Ojeda2015}.  Such stars are often rapidly rotating and very active, especially if the system age is $\sim1$ Gyr \citep{Gaidos2023b}. The $o$ filter but not the $c$ filter includes the H$\alpha$ line that intensifies during flares.  There are also long term (months) variation of $\sim$0.04 mag in the $o$-band (but not $c$-band) which are correlated with PSF FWHM (~3" or more) and are undoubtedly variable contamination from this very red star.

\begin{figure*}
    \centering
    \includegraphics[width=\textwidth]{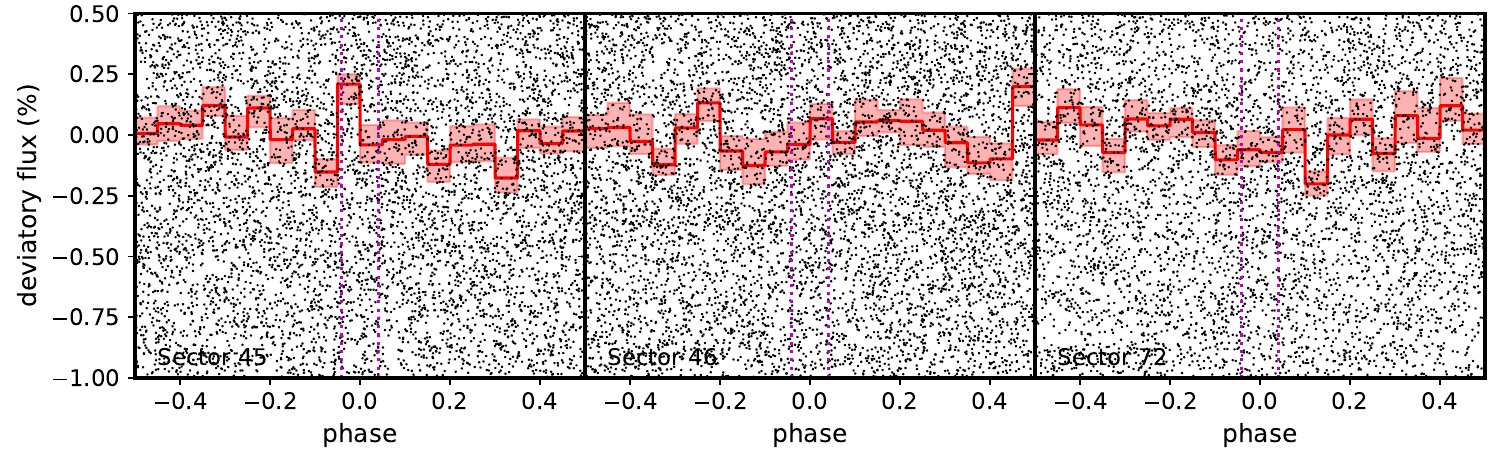}
    \caption{\tess{} lightcurves of K2-22 for Sectors 45, 46, and 72 phased to the orbital period of ``b".  The red curve is the median in 20 bins and the shaded region is the 68\% confidence range based on 1000 re-samples with replacement.  The vertical dotted lines are the predicted phases of ingress and egress based on a 46-min transit duration \citep{Sanchis-Ojeda2015}.  Many individual data points outside the range are not shown.}
    \label{fig:tess}
\end{figure*}
    
\begin{figure*}
    \centering
    \includegraphics[width=0.49\textwidth]{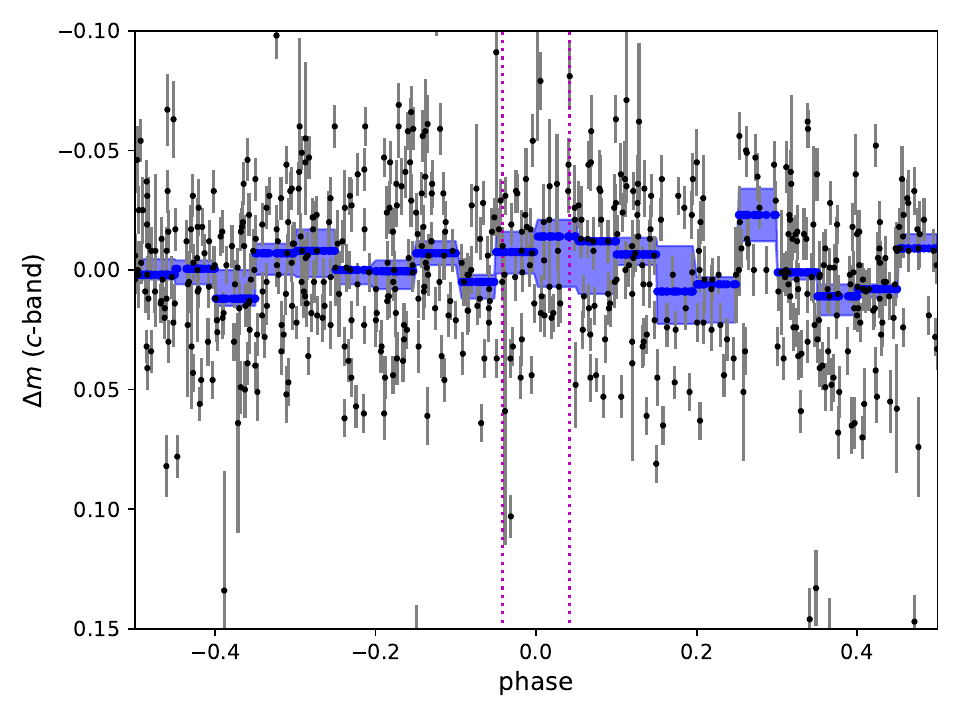}
    \includegraphics[width=0.49\textwidth]{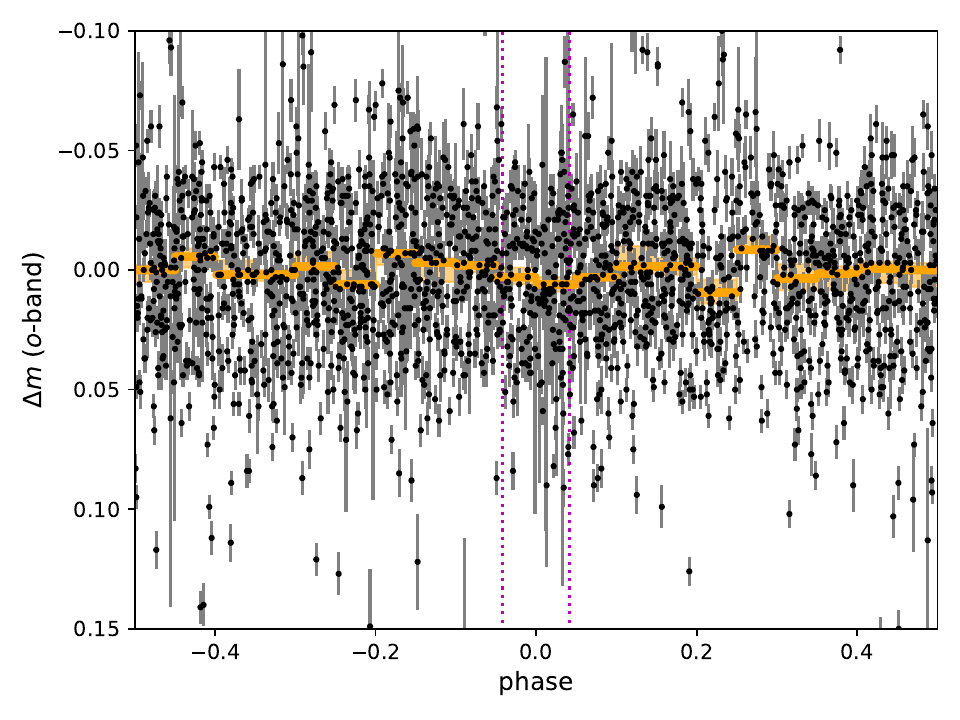}
    \caption{Phased lightcurves of K2-22 from forced photometry in ATLAS images through ``cyan" (left) and ``orange" (right) filters.  The curves are the medians in 20 bins and the shaded regions are the 68\% confidence range based on 1000 samples with replacement.  The vertical dotted lines are the predicted phases of ingress and egress based on a 46-min transit duration \citep{Sanchis-Ojeda2015}.}
    \label{fig:atlas}
\end{figure*}

\section{Comparison of \Angstrom Coefficients to a Mie Scattering Model}
\label{app:scattering}

\begin{figure*}
    \centering
    \includegraphics[width=\columnwidth]{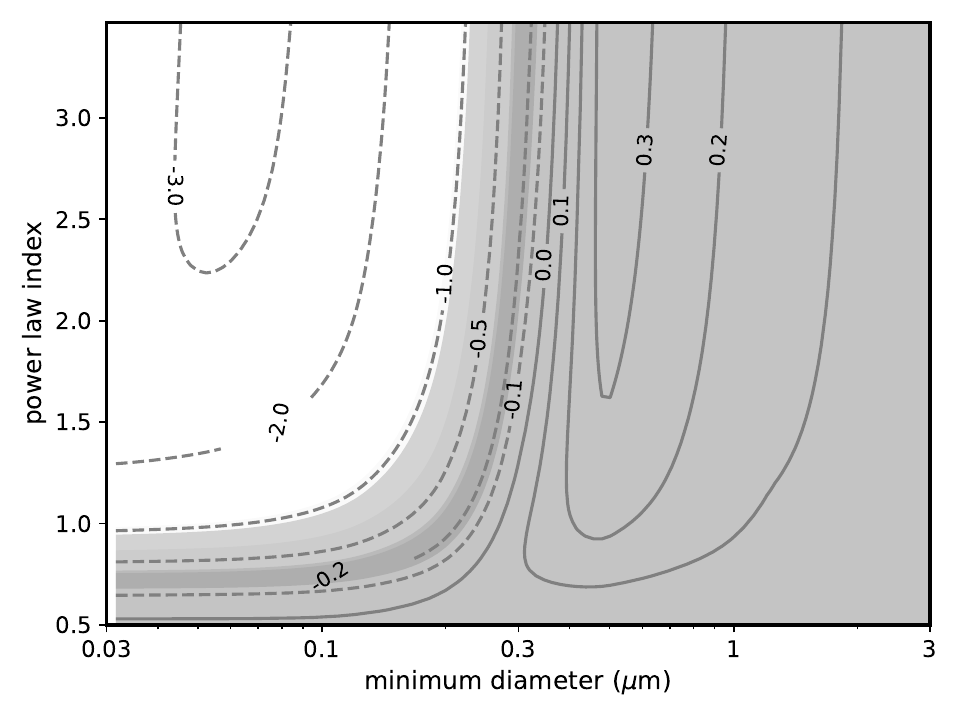}
    \includegraphics[width=\columnwidth]{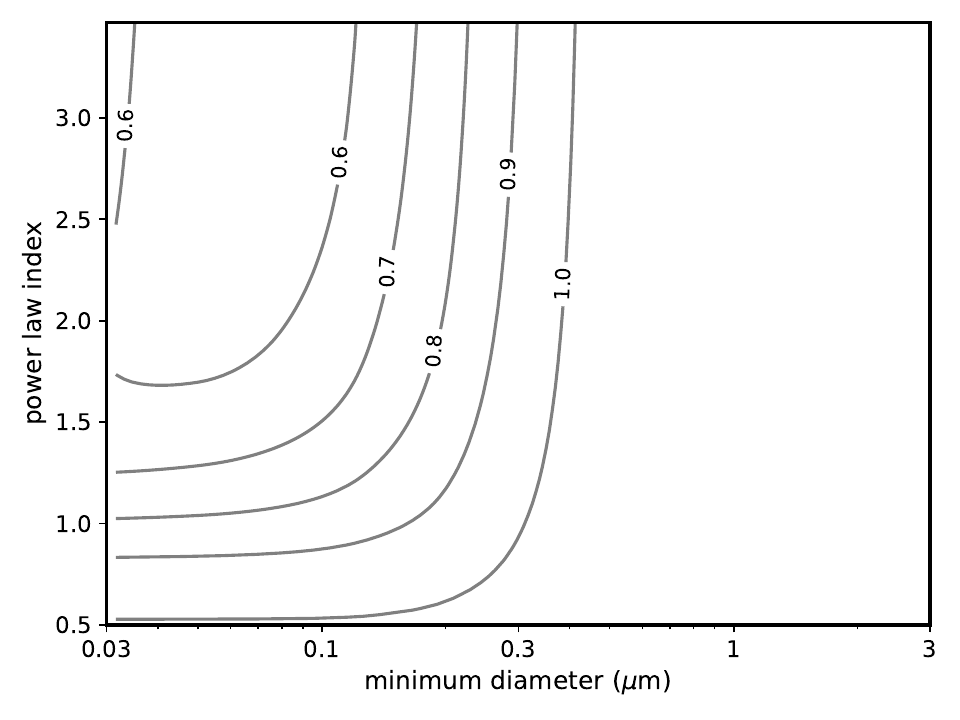}
    \caption{Left (a): Predicted \Angstrom coefficient for $griz$ wavelengths as a function of minimum dust size and power-law index.  Grey regions indicate the range of values derived from each of five transit observations.  Right (b): Predicted ratio of \tess{} to \kepler{} transit depth due to scattering by dust.}
    \label{fig:olivine_angexp}
\end{figure*}

 We compared the estimated \Angstrom coefficients to those predicted by a model of single Mie scattering by spherical dust grains \citep{Budaj2015}.  We adopted a power-law dust size distribution with a minimum size cutoff where the power-law index and cutoff are varied.  We adopted a dwarf K7 spectral type template from \citet{Kesseli2016} for K2-22 and computed single-scattering opacities over the MuSCAT passbands using profiles from the Spanish Virtual Observatory filter service.   We considered several compositions (Mg-Fe olivine, enstatite, forsterite) and found similar results because the opacity at these wavelengths is dominated by scattering.  Figure \ref{fig:olivine_angexp}a show the case for olivine; for non-flat distributions the minimum size is constrained to $\sim0.3$\micron.  Smaller grains are allowed if there is multiple scattering within an optically thick cloud.  We used the same approach to calculate the ratio of the opacities of dust in the \tess{} to \kepler{} passbands.  For grain size distributions that are consistent with the MuSCAT data, that ratio is $\gtrsim0.9$  (Fig. \ref{fig:olivine_angexp}b).

\end{appendix}

\end{document}